\documentclass[reprint,superscriptaddress]{revtex4-2}
\usepackage{amssymb,amsmath,amsthm}
\usepackage{graphicx}
\usepackage{xcolor}
\usepackage{bm,bbm}
\usepackage[bookmarks=false]{hyperref}
\hypersetup{colorlinks=true,citecolor=blue,linkcolor=red,urlcolor=blue,pdfstartview=FitH,bookmarksopen=true}
\usepackage{float}
\usepackage{comment}
\usepackage{mathtools}
\usepackage{enumitem}
\usepackage{txfonts}
\usepackage{algorithm,algpseudocode}

\usepackage{contour}
\usepackage{geometry}
\usepackage{booktabs}
\newcommand{\ket}[1]{\text{$ | #1 \rangle $}}
\newcommand{\bra}[1]{\text{$ \langle #1 | $}}

\newcommand{\tr}{\mathrm{Tr}}

\newcommand{\re}{\mathrm{e}}
\newcommand{\rI}{\mathrm{i}}

\contourlength{0.1pt}

\contourlength{0.1pt}

\newtheoremstyle{note}
  {\topsep/2}               % ABOVE SPACE
  {\topsep/2}            	  % BELOW SPACE
  {}                        % BODY FONT
  {\parindent}              % INDENT
  {\itshape}                % HEAD FONT
  {.---}                    % HEAD PUNCTUATION
  {0pt}                     % HEAD SPACE
  {\thmname{#1}\thmnumber{ \itshape#2}\thmnote{ (#3)}} % CUSTOM-HEAD-SPEC

\theoremstyle{definition}

\theoremstyle{remark}

\usepackage{ulem}

\usepackage{cancel}
\begin{document}
\title{Prescriptive preparation and verification of nonstabilizer states}

\author{Jian Li}
\affiliation{Key Laboratory of Advanced Optoelectronic Quantum Architecture and Measurement (MOE), School of Physics, Beijing Institute of Technology, Beijing 100081, China}

\author{Ye-Chao Liu}
\email{liu@zib.de}
\affiliation{Zuse-Institut Berlin, Takustra{\ss}e 7, 14195 Berlin, Germany}
\affiliation{Key Laboratory of Advanced Optoelectronic Quantum Architecture and Measurement (MOE), School of Physics, Beijing Institute of Technology, Beijing 100081, China}

\author{Xiao-Xiao Chen}
\affiliation{Key Laboratory of Advanced Optoelectronic Quantum Architecture and Measurement (MOE), School of Physics, Beijing Institute of Technology, Beijing 100081, China}

\author{Zhe Meng}
\affiliation{Key Laboratory of Advanced Optoelectronic Quantum Architecture and Measurement (MOE), School of Physics, Beijing Institute of Technology, Beijing 100081, China}

\author{Xing-Yan Fan}
\affiliation{Key Laboratory of Advanced Optoelectronic Quantum Architecture and Measurement (MOE), School of Physics, Beijing Institute of Technology, Beijing 100081, China}

\author{Wen-Hao Wang}
\affiliation{Key Laboratory of Advanced Optoelectronic Quantum Architecture and Measurement (MOE), School of Physics, Beijing Institute of Technology, Beijing 100081, China}

\author{Jie Ma}
\affiliation{Key Laboratory of Advanced Optoelectronic Quantum Architecture and Measurement (MOE), School of Physics, Beijing Institute of Technology, Beijing 100081, China}

\author{An-Ning Zhang}
\email{anningzhang@bit.edu.cn}
\affiliation{Key Laboratory of Advanced Optoelectronic Quantum Architecture and Measurement (MOE), School of Physics, Beijing Institute of Technology, Beijing 100081, China}

\author{Jiangwei Shang}
\email{jiangwei.shang@bit.edu.cn}
\affiliation{Key Laboratory of Advanced Optoelectronic Quantum Architecture and Measurement (MOE), School of Physics, Beijing Institute of Technology, Beijing 100081, China}

\date{July 28, 2026}
%

%%%%%%
\begin{abstract}
High-fidelity quantum state preparation is a central task in quantum information science. 
In practice, it is commonly guided either by full quantum state tomography, which becomes prohibitively resource-intensive as system size grows, or by empirically chosen measurement settings that lack principled optimality. 
Here we show that quantum state verification (QSV) can be elevated from a purely diagnostic tool to a prescriptive framework for quantum state preparation, directly specifying experimentally optimal measurements and quantitative fidelity indicators without full state reconstruction.
We experimentally realize this prescriptive paradigm using a three-qubit nonstabilizer $W$ state and a modified homogeneous QSV protocol. 
The verification measurements not only certify the prepared state with high confidence but also serve as a tomography-free indicator that systematically informs the preparation procedure. 
Using only nine measurement settings and $10^4$ samples, we achieve high-fidelity state preparation consistent with full tomography that requires orders of magnitude more resources. 
Beyond the present implementation, the prescriptive structure of QSV is naturally compatible with closed-loop feedback control, outlining a pathway toward genuine real-time quantum state preparation in future low-latency platforms.
\end{abstract}

\maketitle
%

%%%%%%
\textit{Introduction.---}%
Not only represents the most fundamental feature of quantum physics, entanglement also underpins various significant applications in quantum information science. Over the past decades, substantial efforts have been devoted to generating high-quality entangled states. While the preparation of bipartite entangled states, especially Bell states, is now a mature technique, the realization of multipartite entanglement remains challenging.
Among multipartite entangled states, the Greenberger–Horne–Zeilinger (GHZ) states \cite{GHZ89} are widely studied due to their genuine multipartite entanglement (GME) and stabilizer structure, which allows for efficient simulations.
Meanwhile, $W$ states, and more generally Dicke states \cite{Dicke1954}, are of particular interest owing to their GME properties but lack a stabilizer description. 
These states arise naturally in fermionic systems and play key roles in quantum memories, multiparty quantum networks, universal cloning machines, and precision metrology. More significantly, their robustness against particle loss makes them highly attractive for practical applications.

To characterize entangled states, quantum state tomography (QST) \cite{QSE2004} remains the standard approach. 
Although it can fully reconstruct the density matrix, QST is both time-consuming and computationally intensive \cite{QSTshang}. 
Therefore, substantial efforts have been made to construct more efficient non-tomographic methods \cite{Toth_2005_detecting, Flammia_2011_direct, Aolita_2015_reliable, Hayashi_2015_verifiable, Takeuchi.Morimae2018,PhysRevApplied.23.064005}.
However, many of these approaches lack efficiency or generality for nonstabilizer states. 
Alternatively, quantum state verification (QSV) \cite{Pallister.etal2018} has emerged as a powerful tool for high-precision verification, 
ensuring that a prepared state is sufficiently close to the target state using only local operations and classical communication. 
Up to now, efficient QSV protocols have been developed for various entangled states \cite{Morimae.etal2017, Takeuchi.Morimae2018, Yu.etal2019, Li.etal2019, Wang.Hayashi2019, Zhu.Hayashi2019a, Zhu.Hayashi2019b, Zhu.Hayashi2019c, Zhu.Hayashi2019d, Liu.etal2019b, Li.etal2020b, Dangniam.etal2020, zhang_2019_experimental, jiang_2020_towards, zhang_2020_classical, Li.etal2020a, Liu.etal2020b,Liu.etal2021, Han.etal2021, Zhu.etal2022, ChenHuang.etal2023, Li.Y.etal2021, Yu.etal2022}, including nonstabilizer states \cite{Liu.etal2019b, Li.etal2020a}.
In addition, several QSV experimental demonstrations have been reported for bipartite entangled states \cite{zhang_2019_experimental,jiang_2020_towards,zhang_2020_classical,xia_2022_experimental}.

Despite these advances, previous QSV experiments have been focused exclusively on verifying the correctness of the protocols themselves. 
In all such demonstrations, the entangled states were first carefully prepared, typically using QST, to ensure a high fidelity, and only then subjected to verification. 
As a result, QSV has so far played a passive, post-hoc role, rather than being actively involved in the state preparation process. 
The overall procedure for achieving and certifying high-fidelity entanglement from initially untuned devices remains inefficient and resource-intensive.

\begin{figure}[t]
    \centering
    \includegraphics[width=0.95\linewidth]{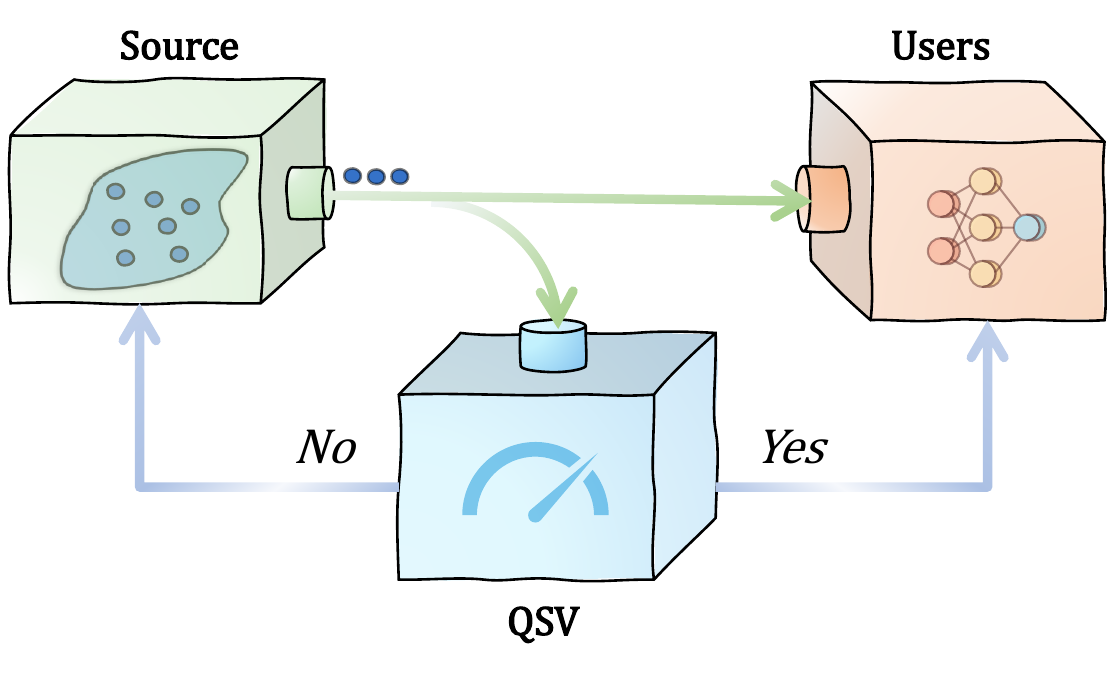}
    \caption{Schematic illustration of prescriptive preparation and verification of entangled states. An entanglement source generates candidate states, which are verified using a quantum state verification protocol based on local measurements. The verification results are used as indicators to adjust the source parameters, enabling fine-tuning of the state preparation. This loop yields high-fidelity entangled states without resorting to full quantum state tomography.
    In principle, the processes can be realized using low-latency control hardware to construct real-time closed-loop optimization.}
    \label{fig:idea}
\end{figure}

In contrast to this conventional post-verification paradigm, we show that QSV can be elevated from a purely diagnostic tool to a prescriptive framework for quantum state preparation, in which the verification protocol itself directly specifies experimentally optimal measurement prescriptions and quantitative fidelity indicators.
In our implementation, this prescriptive QSV framework enables systematic and tomography-free refinement of the preparation procedure by translating verification outcomes into experimentally actionable information; see Fig.~\ref{fig:idea}.
Using only $10^4$ QSV tests, we certify a fidelity of $97.07(\pm 0.26)\%$, which is independently confirmed by QST to be $98.58(\pm 0.12)\%$ with approximately $10^6$ measurements.
Our results constitute the first experimental realization of QSV for nonstabilizer $W$ states, and more significantly, the first demonstration of using QSV to actively assist in entangled state preparation. 
This establishes QSV as a resource-efficient alternative to full tomography for prescriptive quantum state engineering, with natural compatibility with future real-time implementations based on closed-loop feedback control.

%%%%%%
\textit{Theoretical framework.---}%
Consider a quantum device that intends to produce a specific target state \ket{\psi}, but in practice outputs a sequence of independent states ${\sigma_1, \sigma_2, \cdots, \sigma_N}$ over $N$ rounds. 
The goal of QSV is to determine, with confidence at least ${1-\delta}$, whether the device is in an ideal regime where ${\sigma_i=\ket{\psi}\bra{\psi}}$ for all $i$, or in the faulty regime where ${\bra{\psi}\sigma_i\ket{\psi} \leq 1-\epsilon}$ for all $i$.
A QSV protocol can be expressed as a measurement operator ${\Omega = \sum_l p_l\Omega_l}$, where ${\{\Omega_l,\openone-\Omega_l\}}$ are two-outcome measurements and $\{p_l\}$ forms a probability distribution. 
The measurements are designed to satisfy ${\bra{\psi}\Omega_l\ket{\psi}=1}$, ensuring the target state \ket{\psi} always passes the test.
Therefore, the worst-case probability for a faulty state to pass the test is ${\max_{\bra{\psi}\sigma\ket{\psi} \leq 1-\epsilon}\tr(\Omega\sigma)=1-\epsilon\nu}$, where ${\nu=1-\lambda}$ is the spectral gap between the largest and the second-largest eigenvalues of $\Omega$.
In order to verify a target state within infidelity $\epsilon$ and significance level $\delta$, the required number of tests is \cite{Pallister.etal2018}
\begin{equation}\label{eq:QSV}
    N = \ln\delta^{-1}/\ln(1-\epsilon\nu)^{-1} \approx \frac{1}{\nu}\epsilon^{-1}\ln\delta^{-1}\,.
\end{equation}

One significant advantage of the QSV strategy lies in its sample complexity scaling as $O(\epsilon^{-1})$, approaching the Heisenberg limit. 
This makes it particularly suitable for high-precision scenarios, outperforming quantum tomography and other non-tomographic methods. 
Meanwhile, many entangled states can be verified using only local operations with resource overheads that scales as polynomial, linear, or even constantly with the system size.
And the employment of classical communication, i.e., adaptive measurements, can further reduce the overhead \cite{Yu.etal2019, Li.etal2019, Wang.Hayashi2019}.

\begin{figure}[t]
    \centering
    \includegraphics[width=0.95\linewidth]{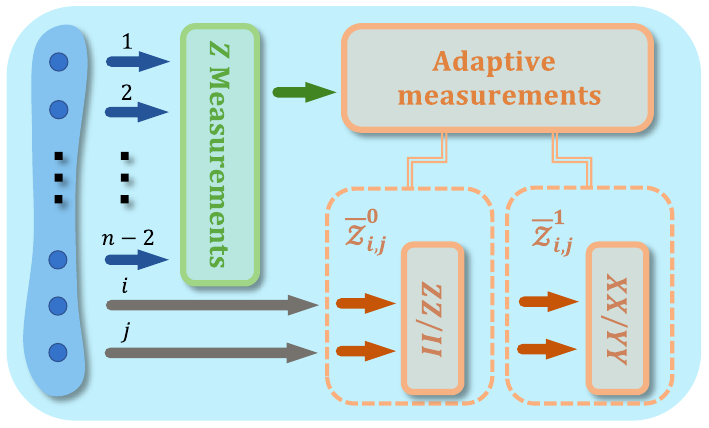}
    \caption{Verification protocol for an $n$-qubit nonstabilizer state $W_{n}$ via two-step adaptive measurements. For any two qubits $i$ and $j$ selected a priori, the measurement outcomes of the other ${n-2}$ qubits determine which measurements to conduct on them. The notation $\mathcal{\bar{Z} }_{i,j}^{k}$ denotes that $k$ excitations are detected if one performs Pauli-$Z$ measurements on all qubits except for $i$ and $j$.}
    \label{fig:1}
\end{figure}

For three-qubit systems, the GHZ and $W$ states are the only two inequivalent classes of GME states.
While GHZ states admit a stabilizer description that facilitates efficient verification using local Pauli measurements \cite{Flammia_2011_direct, daSilva.etal2011, Aaronson2018, Takeuchi.Morimae2018, Pallister.etal2018, Dangniam.etal2020, Kliesch.etal2021_theory}, the verification of nonstabilizer states such as $W$ states is significantly harder. In fact, the sample complexity is directly linked to the state's nonstabilizerness \cite{Leone.etal2023_nonstabilizerness}, making generic approaches exponentially inefficient.
Luckily, this challenge was tackled in Ref.~\cite{Liu.etal2019b}, which introduced an efficient QSV protocol for $W$ and Dicke states using only local operations and classical communication.
For an $n$-qubit $W$ state
\begin{equation}
    W_n = \frac{1}{\sqrt{n}}\bigl(\ket{10\cdots0}+\ket{01\cdots0}+\cdots+\ket{00\cdots1}\bigr)\,,
\end{equation}
the verification protocol reads
\begin{equation}
    \begin{aligned}
        &\Omega \left (W_n \right ) 
        =\frac{2}{n(n-1)}\sum_{i<j} \Omega_{i,j}^{\to}  \,,\\
        &\Omega_{i,j}^{\to} 
        = \mathcal{\bar{Z} }_{i,j}^{1}\bigl(Z_i^{+}Z_j^{+}\bigr) + \mathcal{\bar{Z} }_{i,j}^{0}\bigl(XX\bigr)_{i,j}^{+} \,.
    \end{aligned}
\end{equation}
As illustrated in Fig.~\ref{fig:1}, $\Omega_{i,j}^{\to}$ is the one-way adaptive measurement.
The measurement $\mathcal{\bar{Z} }_{i,j}^{k}$ proceeds by performing $Z$ measurements on all qubits except for $i$ and $j$ to count the number of excitations.
Depending on the outcome ${k\in\{0,1\}}$, one performs either a local Pauli projection $Z_i^{+}Z_j^{+}$ ($k=1$) or $\bigl(XX\bigr)_{i,j}^{+}$ ($k=0$).
The sample complexity of the protocol is given by
\begin{equation}
    N = \Biggl\{ 
    \begin{aligned}
        &3\epsilon^{-1}\ln\delta^{-1}       &\text{for}\quad n= 3\,,\\
        &(n-1)\epsilon^{-1}\ln\delta^{-1}   &\text{for}\quad n\geq 4\,.
    \end{aligned}
    \,
\end{equation}

\begin{figure*}[t]
    \centering\includegraphics[width=0.95\linewidth]{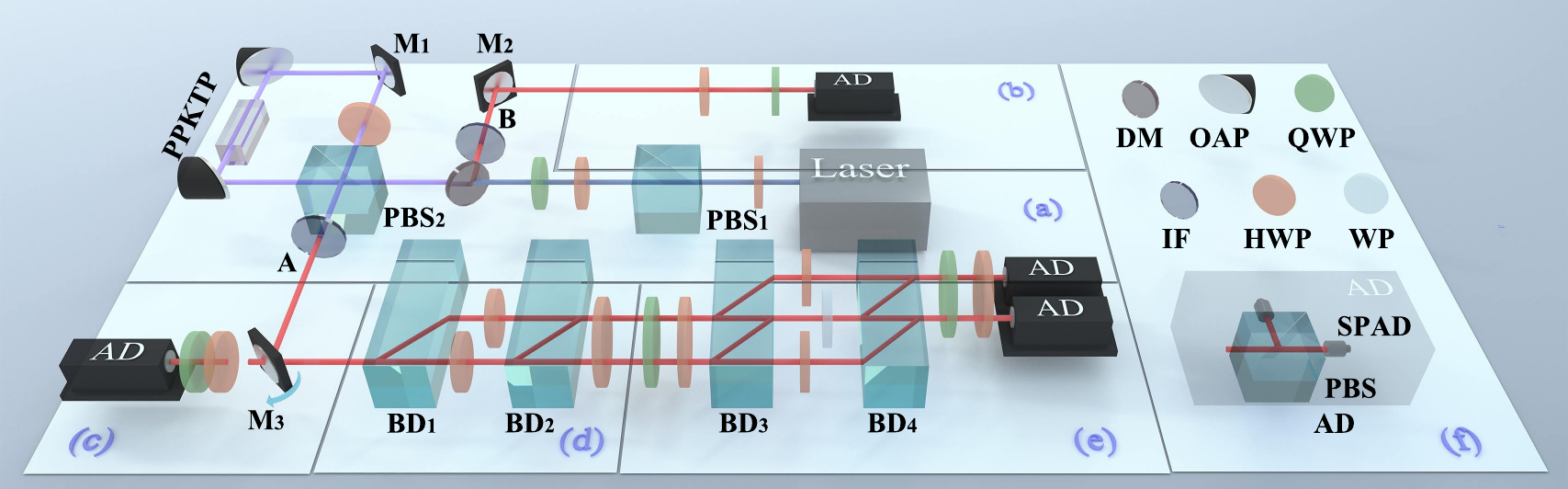}
    \caption{Experimental setup for the prescriptive preparation and verification of two-qubit states and a three-qubit nonstabilizer $W_{3}$ state. The two-photon entanglement source (a) is realized through an interferometer and spontaneous parametric down-conversion process (SPDC) in a periodically poled potassium titanyl phosphate (PPKTP) crystal. Mirror M$_{3}$ can be rotated to different angles. The setup in (b) shows the measurement apparatus for photon B, which remains unchanged. For two-qubit quantum states, photon A can be measured directly in (c). The nonstabilizer state is prepared in (d) and the verification is utilized in (e). The devices used in our experiment are illustrated in (f). Each analysis device (AD) consists of a polarizing beam splitter (PBS) and two couplers connected to single-photon avalanche diodes (SPADs), allowing the measurement of different eigenstates; DM: dichroic mirror; OAP: off-axis parabolic mirror; IF: 10nm bandwidth interference filter; HWP: half-wave plate; QWP: quarter-wave plate; WP: window plate; BD: beam displacer.
 }
    \label{expset}
\end{figure*}

Beyond their high efficiency in sample complexity, QSV protocols are also remarkably hardware-efficient with respect to the number of required measurement settings~\cite{Li.Y.etal2021}.
This feature makes QSV highly promising for quantum state preparation.
Typically, the tuning process of state preparation relies on an indicator derived from quantum state tomography (or, for Bell states, a Bell-inequality test), which demands an exponential number of measurement bases with the system size~\cite{meng.etal2023}.
Although strategies with fewer bases exist, they are typically empirical and thus unsuitable for general cases~\cite{neumann_continuous_2022, yin_polarization_2025}.
QSV protocols overcome this limitation by offering hardware-saving measurement schemes and providing a natural fidelity-lower-bound indicator applicable to arbitrary states, making them an ideal tool for efficient quantum state preparation, particularly for the nonstabilizer states studied in this work.
Moreover, for homogeneous QSV protocols, the fidelity can be obtained exactly rather than as a lower bound \cite{Zhu.Hayashi2019c,Zhu.Hayashi2019d}.
To conclude, QSV-based indicators that are simultaneously efficient in both sample complexity and measurement settings provide a prescriptive alternative to empirical, heuristic measurement choices.
From a broader perspective, the QSV-based framework is naturally compatible with closed-loop feedback optimization of state preparation, and thus offers a pathway toward future real-time-capable implementations with sufficiently low-latency hardware \cite{rubenok_real-world_2013, wengerowsky_entanglement_2019, neumann_continuous_2022, meng.etal2023, craddock_automated_2024, yin_polarization_2025}.

%%%%%%
\textit{Experimental settings.---}%
In our experiment, we focus on a three-qubit $W_3$ state and adopt a modified QSV protocol \cite{Liu.etal2023_effcient}
\begin{equation}
    \label{eq:OmegaHom}
    \begin{aligned}
        &\Omega_{\rm Hom}\left ( W_{3} \right ) 
        =\sum_{k=1}^{3}\frac{1}{3}\mathcal{P}_{k}\bigl\{ \Omega^{\to} \bigr\}  \,,\\
        &\Omega^{\to} 
        = P_{Z}^{+}\left [ \frac{1}{2}P_{XX}^{+}+ \frac{1}{2}P_{YY}^{+} \right ]
        +P_{Z}^{-}\left [ \frac{1}{2}\mathbb{II}+ \frac{1}{2}P_{Z}^{+}P_{Z}^{+}\right ].
    \end{aligned}
\end{equation}
Here, $\Omega^{\to}$ is the one-way adaptive measurement from the first qubit to the others, and ${\mathcal{P}_k\,(k=1,2,3)}$ denotes all permutations of the three qubits. 
The measurement $P_O^{+(-)}$ is the projection onto the positive (negative) eigenspace of the local operator $O$, constructed from Pauli operators $X, Y, Z$. 

This modified version has three advantages.
First, it improves the verification efficiency for the three-qubit $W_3$ state, i.e.,
\begin{equation}
    N = 2\epsilon^{-1}\ln\delta^{-1}\,,
\end{equation}
reducing the sample complexity by one-third as compared with the original protocol in Ref.~\cite{Liu.etal2019b}.
Second, it represents a homogeneous protocol \cite{Zhu.Hayashi2019c,Zhu.Hayashi2019d}, making it suitable for practical scenarios, especially for our prescriptive preparation of entangled states as discussed above.
Finally, the improvement requires only additional local Pauli-$Y$ projections, which constitute a standard and highly accurate experimental setup, fully consistent with our key motivation for high-precision state preparation.

The schematic diagram in Fig.~\ref{expset} illustrates the experimental demonstration of the modified QSV protocol for three-qubit  $W_3$ state. 
The two-qubit entanglement source, as shown in Fig.~\ref{expset}(a),  is used to generate a two-photon polarization entangled state. To adjust the intensity of the 405nm continuous-wave pump light, a half-wave plate (HWP) and a polarizing beam splitter (PBS) PBS$_1$ are utilized.
Subsequently, the pump light is split into two paths by a dual-wavelength PBS$_2$. The polarization of the pump light is adjusted to horizontal polarization using a HWP. The off-axis parabolic mirrors (OAPs) focus the pump light onto a periodically poled potassium titanyl phosphate (PPKTP) crystal to generate type-II spontaneous parametric down-conversion (SPDC) photon pairs at 810nm. After being collimated by OAPs, the photon pairs meet again at PBS$_2$. Assisted by PBS$_2$ and a dichroic mirror, these photon pairs from the SPDC process will be separated onto paths A and B. By adjusting the angles of the HWP and QWP after PBS$_1$, we can prepare two-qubit states in the form of $\alpha\ket{01}_{\rm AB} + \re^{\rI\varphi}\beta\ket{10}_{\rm AB}$. The logical qubits are encoded as $\ket{0} = \ket{H}$ and $\ket{1} = \ket{V}$, where $H$ ($V$) denotes the horizontal (vertical) polarization. More details about this entangled source can be found in Ref.~\cite{meng.etal2023}. 
We have assessed the performance of the entanglement source by preparing a two-qubit Bell state $\frac{1}{\sqrt{2}}\bigl(\ket{01}_{\rm AB} + \ket{10}_{\rm AB}\bigr)$. 
The Clauser-Horne-Shimony-Holt (CHSH) inequality \cite{clauserProposedExperimentTest1969} is then tested, yielding a non-locality violation with a value of $2.8088\pm0.0045$, confirming the proper functionality. See more details in Appendix~A . 

\begin{figure*}[t]
    \centering
    \includegraphics[width=0.95\linewidth]{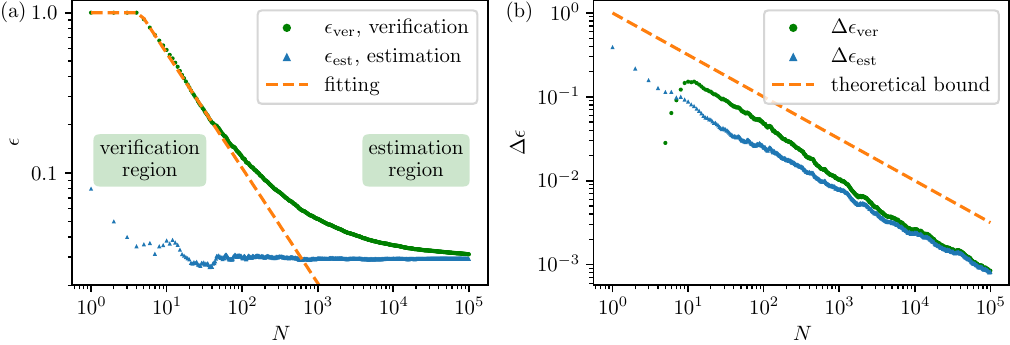}
    \caption{(a) Verified infidelity $\epsilon_{\rm ver}$ (green dots, with the significance level ${\delta < 0.05}$) and directly estimated infidelity $\epsilon_{\rm est}$ (blue triangles) as functions of the number of tests $N$. All values are derived from $100$ experimental trials. The estimated infidelity is consistently lower than the verified bound, as expected. The dashed line indicates the numerical fit for the scaling behavior with $O(\epsilon^{-1.39})$. (b) Statistical deviation of the estimated average fidelity of verification (green dots) and estimation (blue triangles) compared with the theoretical upper bound from the QSV protocol (orange dashed line).} 
    \label{fig:results}
\end{figure*}
To achieve the one-way adaptive measurement as in Eq.~\eqref{eq:OmegaHom}, we initially prepare entangled states of two qubits using polarization degrees of freedom (DOFs). 
They are first prepared to achieve high quality by tuning devices via the assistance of QSV and are meanwhile certified with the QSV protocol as well.
The detailed discussion is reported in Appendix~B, including the compensation algorithm for tuning.
An independent parallel process was performed via the assistance of QST to preliminarily benchmark the state preparation capabilities of QSV, and a comparison is presented in Appendix~C .
Arbitrary Pauli projections required for QSV can be implemented with the coincidence counts being recorded. The QSV and QST are performed on these two-qubit states using the analysis setups shown in Figs.~\ref{expset}(b) and \ref{expset}(c).

Next, by introducing a path qubit through a DOF multiplexing method, we generate the nonstabilizer $W_{3}$ state on which the QSV protocol in Eq.~\eqref{eq:OmegaHom} is successfully implemented. Specifically, by rotating M$_3$, photon A is directed into a polarization-path DOF encoding device, which consists of three HWPs and two beam displacers (BDs), labeled BD$_1$ and BD$_2$. This device, as shown in Fig.~\ref{expset}(d), introduces the path DOF for photon A. After passing through the encoding setup, the state is transformed into $\frac{1}{\sqrt{3}}(\ket{001}+\ket{100}+\ket{010})$, which corresponds to a standard three-qubit $W_{3}$ state in the Hilbert space $\mathcal{H}_{\rm A_{pol}} \otimes \mathcal{H}_{\rm A_{pat}} \otimes \mathcal{H}_{\rm B_{pol}}$. The ultra logical qubit is encoded as $\ket{0} = \ket{U}$ and $\ket{1} = \ket{L}$, where $U$ ($L$) denotes the upper (lower) path of photon A.
The prepared $W_{3}$ state is subsequently sent into the three-qubit state analyzer, as shown in Figs.~\ref{expset}(b) and \ref{expset}(e). During the experiment, the angles of the QWPs and HWPs in (b) and (e) are adjusted by QSV-based prescriptive framework to implement the measurements described by Eq.~(\ref{eq:OmegaHom}). 
More discussions on the tuning process are included in Appendix~D .
In addition, QST is performed on the experimental state. A mutual validation between QSV and QST is also carried out, specifically tailored for the analysis of the $W_{3}$ state; see Appendix~E for details .

%%%%%%
\textit{Results.---}%
Because of unavoidable systematic error, the experimentally prepared states inevitably deviate from the ideal target. 
This can lead to an occasional failure of verification with a certain number of measurements. 
In such a scenario, the QSV strategy can always be converted to a more practical model to certify a lower bound of the average fidelity of the device's output \cite{Yu.etal2022}, which has been adopted in several recent experiments \cite{zhang_2019_experimental,jiang_2020_towards,zhang_2020_classical,xia_2022_experimental}.
To proceed, a hypothesis test can be formulated with the null and alternative hypotheses being
\begin{equation}\label{eq:htest}
    \begin{aligned}
        H_0: \frac{1}{N}\sum_{i=1}^{N}\bra{\psi}\sigma_i\ket{\psi} \leq 1-\epsilon\,,\\
        H_1: \frac{1}{N}\sum_{i=1}^{N}\bra{\psi}\sigma_i\ket{\psi} > 1-\epsilon\,.
    \end{aligned}
\end{equation}

Hence, with the same experimental settings, i.e., $\{\Omega_l,\openone-\Omega_l\}$, one can get a frequency ${f=t/N}$ with $t$ passing instances out of $N$ tests.
The null hypothesis can then be rejected and conclude that the average fidelity is larger than $1-\epsilon$ if
\begin{equation}
    f > 1 - \epsilon\nu\,,
\end{equation}
with the significance level given by the Chernoff-Hoeffding theorem \cite{Hoeffding1994}
\begin{equation}\label{eq:QFE}
    \delta < {\rm e}^{-D[f||(1-\epsilon\nu)]N}\,,
\end{equation}
where ${D(x\|y)\equiv x\log\frac{x}{y}+(1-x)\log\frac{1-x}{1-y}}$
is the Kullback–Leibler divergence. 
Note that if all the tests had passed, namely $f=1$, Eq.~\eqref{eq:QFE} reduces to Eq.~\eqref{eq:QSV}.

In Fig.~\ref{fig:results}(a), we illustrate the minimal infidelity $\epsilon_{\rm ver}$ verified at a significance level ${\delta < 5\%}$ based on the experimental data. 
Because of statistical fluctuations in a single run of QSV, all reported results are averaged over $100$ independent trials.
As expected, in general, the infidelity from practical scenarios based on Eq.~\eqref{eq:QFE} deviates from the ideal QSV scaling. 
Nevertheless, if one focuses on the low-sample region, the advantage of the QSV measurement protocol, which scales as $O(\epsilon^{-1})$ with respect to the number of samples, is partially preserved as $O(\epsilon^{-1.39})$ based on the fitting to the experimental data.
In this region, the verification task can actually be executed well in our experiment.
For instance, $20$ tests can guarantee a fidelity greater than $65\%$, and $100$ tests are enough to verify a fidelity exceeding $88\%$.

Beyond verification, estimating the average fidelity of the prepared state is essential for evaluating the performance of quantum state preparation. In our experiment, the homogeneous structure of the QSV operator allows a direct estimation of the average fidelity \cite{Zhu.Hayashi2019d}, i.e.,
\begin{equation}
    \epsilon = \frac{1 - f}{\nu}\,,\quad F = 1 - \epsilon = \frac{f - (1-\nu)}{\nu}\,.
\end{equation}
The standard deviation is given by
\begin{equation}
    \Delta \epsilon = \Delta F = \frac{1}{\nu}\sqrt{\frac{f(1-f)}{N}} \leq \frac{1}{2\nu\sqrt{N}}\,.
\end{equation}
The estimated infidelity $\epsilon_{\rm est}$ is compared with the certified upper bound $\epsilon_{\rm ver}$ in Fig.~\ref{fig:results}(a). As expected, the estimated infidelity is consistently lower, reflecting its role as a mean value rather than a guaranteed bound.
Figure~\ref{fig:results}(b) shows the standard deviation of the estimation, which is below the theoretical upper bound from the QSV protocol. For the region of ${N \geq 1000}$, the deviation of the estimation can be controlled within $1\%$.

To conclude, the QSV protocol implemented in our experiment serves a dual purpose:
(1) with only a few dozen tests, it can rapidly certify a fidelity lower bound, and
(2) with a few thousand tests, it enables the accurate estimation of the average fidelity with small statistical deviation.
Both functions require significantly fewer measurements than full QST.
For comparison, a full QST procedure was also performed on the same quantum device. 
The fidelity obtained via QST is $98.58(\pm 0.12)\%$, which is consistent with the QSV estimation of $97.07(\pm 0.26)\%$. An independent parallel process using QST was performed to preliminarily benchmark the state preparation capability of QSV for the $W_3$ state. A detailed comparison can be found in Appendix~E .
However, the resource consumption differs drastically: QST requires $64$ measurement settings and approximately $10^6$ measurements in total, while our QSV protocol needs only nine measurement settings and $10^4$ measurements to achieve the reported results.
This significant reduction in resources highlights the practical advantage of QSV for efficient and scalable quantum state engineering.
The advantage is also confirmed by the results of two-qubit states as well, which are reported in Appendix~B, then compared in Appendix~C.

%%%%%%
\textit{Summary.---}%
We have experimentally demonstrated a high-efficiency QSV protocol for verifying the nonstabilizer three-qubit $W_3$ state, using a modified homogeneous strategy. 
Crucially, in contrast to its conventional use as a purely diagnostic tool, the QSV protocol in our implementation can be elevated to a prescriptive framework for quantum state preparation by providing quantitative fidelity indicators for target states with minimal measurement resources. 
This dual functionality highlights the practical utility of QSV beyond its conventional role.

These breakthroughs are enabled by the homogeneous structure of the QSV operator, which supports both rigorous verification of the fidelity lower bounds and direct estimation of the average fidelity. As compared with full QST, our QSV-based approach requires significantly fewer measurement settings and samples, yet achieves consistent and reliable estimation of the fidelity.
These features position QSV as a powerful and resource-efficient tool for both quantum state verification and high-precision quantum state engineering in near-term quantum experiments.
Although the present experiment focuses on demonstrating this prescriptive capability rather than low-latency control, the same framework is in principle compatible with closed-loop feedback control and thus holds promise for genuine real-time tasks in future high-speed implementations, including fast state preparation, long-term quantum state stabilization and entanglement distribution.

\acknowledgments
This work was supported by the National Natural Science Foundation of China (Grants No.~92365115, No.~12474480, No.~92265115, and No.~12175014), the National Key R\&D Program of China (Grant No.~2022YFA1404900), and the Beijing Natural Science Foundation (Grant No.~1262037). Y.-C. Liu is also supported by the DFG Cluster of Excellence MATH+ (EXC-2046/1, Project No.~390685689) funded by the Deutsche Forschungsgemeinschaft (DFG).

\appendix
\onecolumngrid

\section{Characterization of the two-qubit state source via CHSH inequality}\label{app:source}

In the following Appendices, we provide additional details of the experimental implementation and analysis tools.
Appendix~\ref{app:source} characterizes the quality of the entangled photon source.
Appendix~\ref{app:2qubit} illustrates the feedback-control tuning procedure for two-qubit states based on QSV, which enables the preparation of high-quality entangled states.
The comparison between two-qubit states optimized via QST and QSV, along with the corresponding experimental results, is presented in Appendix~\ref{app:2qubitQST}.
For completeness, Appendix~\ref{app:W} summarizes the main-text results on the preparation and verification of $W$ states and provides a more detailed discussion of the feedback-control tuning process.
Finally, Appendix~\ref{app:WQST} compares the performance of different optimization indicators based on QST and QSV in the preparation of $W$ states.

To evaluate the performance of the two-qubit entanglement source, we generate a maximally entangled Bell state of the form $\frac{1}{\sqrt{2}}\bigl(\ket{01}_{\rm AB} + \ket{10}_{\rm AB}\bigr)$. This state serves as a benchmark for characterizing the nonlocal correlations between the two photons. To quantify the degree of quantum nonlocality, we calculate the $S$-parameter using the CHSH inequality \cite{clauserProposedExperimentTest1969}.
The CHSH inequality provides a criterion for ruling out local hidden variable theories, and a violation of the classical bound ${S \leq 2}$ certifies the presence of quantum entanglement. In our experiment, the coincidence counts are measured under $16$ different polarization settings required for the CHSH analysis. 

\begin{table}[htbp]
     \centering
     \renewcommand{\arraystretch}{1.1} %
     \setlength{\tabcolsep}{10pt} %
     \begin{tabular}{ccccc}
       \hline
       \hline
        & $22.5^{\circ}$ & $67.5^{\circ}$ & $112.5^{\circ}$ &$157.5^{\circ}$ \\
        \hline
        $ 0^{\circ}$ & 21164 & 119638 & 117702 & 18910\\
         \hline
         $45^{\circ}$& 116151 & 114421 & 20071 & 21873\\
         \hline
        $90^{\circ}$& 115494 & 19061 & 20620 & 114872\\
        \hline
       $ 135^{\circ}$& 19036 & 21698 &116486  &113688 \\
        \hline
     \end{tabular}
     \caption{Coincidence counts for different measurement settings in the CHSH inequality test. Each row corresponds to a polarization measurement basis for photon~A, and each column corresponds to a polarization measurement basis for photon~B.}
     \label{polcounts}
 \end{table}

The measured coincidence counts for the various polarization basis combinations are presented in Table~\ref{polcounts}. From these measurements, we compute and obtain an $S$-value being ${2.8088\pm0.0045}$, which significantly exceeds the classical threshold. Notably, this result corresponds to a violation of the CHSH inequality by 179 standard deviations, strongly confirming the presence of high-quality quantum entanglement. This finding demonstrates that the two-qubit state generated by our setup is suitable for applications requiring reliable and high-fidelity entangled photon pairs.
\section{Preparation and verification of two-qubit states via QSV strategies}\label{app:2qubit}
To preliminarily demonstrate the feasibility of using QSV as an indicator for state preparation, we experimentally prepare a series of two-qubit polarization-entangled states of the form  
\begin{equation}\label{eq:twoqubittheta}
\ket{\psi(\theta)} = \sin\theta \ket{01} + \cos\theta \ket{10}\,,
\end{equation}
where $\theta$ is chosen such that $\sin\theta \in \left\{ 0, \frac{1}{\sqrt{3}}, \frac{1}{\sqrt{2}}, \frac{\sqrt{2}}{\sqrt{3}}, 1 \right\}$.

\subsection{Quantum state verification for arbitrary two-qubit states}

For arbitrary two-qubit states, we introduce the optimal non-adaptive QSV protocol proposed in Ref.~\cite{Pallister.etal2018}, i.e.,  
\begin{align}\label{app_eq:QSV2}
\Omega_{\text{opt}} &= \alpha(\theta) P^-_{ZZ} 
+ \frac{1 - \alpha(\theta)}{3} \sum_{k=1}^3 \left[\openone - \ket{\phi_k}\bra{\phi_k} \right],
\end{align}
where $P^-_{ZZ}$ denotes the projector onto the negative eigenspace of the Pauli operator $ZZ$, and the weight $\alpha(\theta)$ is defined as  
\begin{equation}
\alpha(\theta) = \frac{2 - \sin(2\theta)}{4 + \sin(2\theta)}\,. 
\end{equation}
The projection basis $\ket{\phi_k}$ used in the protocol is given by
\begin{align}
\ket{\phi_1} &= \left( \frac{1}{\sqrt{1+\tan\theta}} \ket{0} + \frac{\re^{\frac{2\pi \rI}{3}}}{\sqrt{1+\cot\theta}} \ket{1} \right) 
\otimes \left( \frac{1}{\sqrt{1+\cot\theta}} \ket{0} + \frac{\re^{-\frac{\pi \rI}{3}}}{\sqrt{1+\tan\theta}} \ket{1} \right), \\
\ket{\phi_2} &= \left( \frac{1}{\sqrt{1+\tan\theta}} \ket{0} + \frac{\re^{\frac{4\pi \rI}{3}}}{\sqrt{1+\cot\theta}} \ket{1} \right) 
\otimes \left( \frac{1}{\sqrt{1+\cot\theta}} \ket{0} + \frac{\re^{-\frac{5\pi \rI}{3}}}{\sqrt{1+\tan\theta}} \ket{1} \right), \\
\ket{\phi_3} &= \left( \frac{1}{\sqrt{1+\tan\theta}} \ket{0} + \frac{1}{\sqrt{1+\cot\theta}} \ket{1} \right) 
\otimes \left( \frac{1}{\sqrt{1+\cot\theta}} \ket{0} - \frac{1}{\sqrt{1+\tan\theta}} \ket{1} \right).
\end{align}
The number of measurements required to verify the target state within infidelity $\epsilon$ and significance level $\delta$ scales as follows: 
\begin{equation}
\label{eq:twoqubitperf}
N_{\text{opt}} \approx (2 + \sin\theta \cos\theta)\, \epsilon^{-1} \ln \delta^{-1}.
\end{equation}

\subsection{State preparation with polarization compensation algorithm}
In our experiments, the two-qubit states are prepared using the polarization degrees of freedom.
The tuning procedure was implemented with an AI-based polarization compensation algorithm employing a gradient-descent method, which was proposed in our previous work of Ref.~\cite{meng.etal2023}; see below.
\begin{center}
\begin{minipage}{0.65\linewidth}
\begin{algorithm}[H]
  \caption{Automatic polarization compensation \cite{meng.etal2023}}
  \label{alg:feedback}
  \begin{algorithmic}[1]
    \Require Current wave-plate angles $\theta_i$ ($i=1,2,\cdots$), objective function $F(\theta_1, \theta_2, \cdots)$
    \For{$i=1,2,\cdots$}
      \State $F_i = F(\theta_1,\cdots,\theta_i+\varepsilon,\cdots)$
      \State $g_i = (F_i-F)/\varepsilon$                                \Comment{Gradient}
      \State $G = G + g_i^2$                                            \Comment{Accumulated gradient}
      \State $\alpha = \alpha_0/(\sqrt{G}+\eta)$                 \Comment{Adaptive learning rate}
      \State $\theta_i = \theta_i + \alpha \cdot g_i$                     \Comment{Update}
    \EndFor
    \Ensure New wave-plate angles $\theta_i$ ($i=1,2,\cdots$)
  \end{algorithmic}
\end{algorithm}
\end{minipage}
\end{center}
\vspace{1em}

In Ref.~\cite{meng.etal2023}, the algorithm was applied to optimize the two-qubit entanglement source, where the objective function was chosen as the $S$-parameter of the CHSH inequality~\cite{clauserProposedExperimentTest1969}.
This choice, however, is comparable to full state tomography in terms of experimental complexity.
In the present work, instead we adopt an objective function from the QSV protocol, which provides a fidelity lower bound with both efficient sample complexity and hardware-saving measurements.
In our implementation, we directly use the passing frequency as the objective function, since it can be obtained directly from measurement outcomes and is monotonically related to the state fidelity.
Although fidelity (or its lower bound) would be a more accurate indicator, this choice enables a fast indicator without additional post-processing.

In our implementation, each evaluation of the objective function
uses 1000 samples for both the QSV- and QST-based schemes. In each tuning iteration, a baseline objective value
$F(\theta_1,\theta_2,\cdots)$ is first evaluated at the current
wave-plate setting, and the perturbed objective values for the
adjustable wave-plate angles are then evaluated relative to this same
baseline. The wave-plate angles are updated simultaneously only after all gradient components have been obtained, and the updated angles then define the baseline for the next tuning iteration. The compensation routine is run for 30 tuning iterations,
which is already sufficient for the optimization to reach an
essentially converged regime. Here $\varepsilon$ denotes the
finite-difference perturbation, while $\eta$ is introduced as a small
numerical-stability term in the adaptive update rule. For simplicity,
we choose $\eta=\varepsilon$ in our implementation.

\begin{figure*}[tb]
    \centering
    \includegraphics[width=0.95\linewidth]{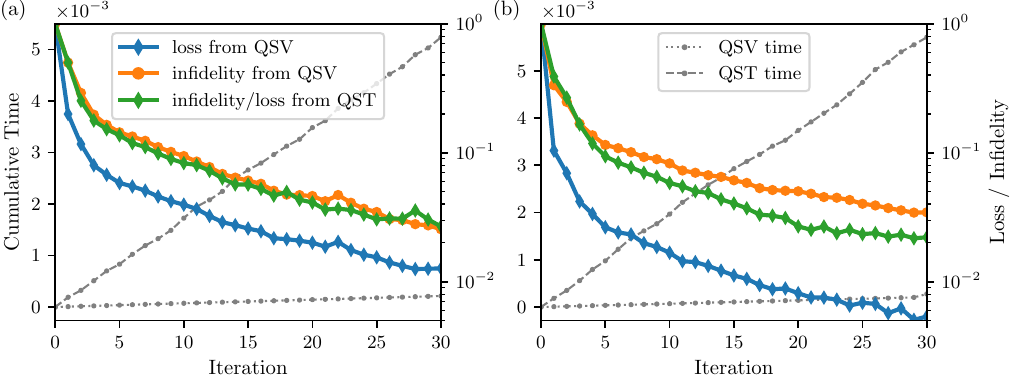}
    \caption{Simulation of prescriptive quantum-state preparation using the polarization-compensation feedback algorithm for (a) $\frac{1}{\sqrt{2}}(\ket{01}+\ket{10})$ and (b) $\frac{1}{\sqrt{3}}\ket{01}+\frac{\sqrt{2}}{\sqrt{3}}\ket{10}$.
    The post-processing time of QSV-based method (gray dotted line) is significantly cheaper than that of QST (gray dashed line).
    The losses evaluated by different objective functions are compared: the passing frequency from QSV (blue diamonds) and the fidelity from full state tomography (green diamonds).
    The infidelity trajectory obtained from the QSV-based method is also shown (orange circles), which is comparable to QST.
    All results are over 100 simulation rounds, and each data point is computed from a 1000-sample measurement.
    }
    \label{fig_app:sim}
\end{figure*}

The effectiveness of the QSV-based prescriptive preparation scheme is demonstrated in Fig.~\ref{fig_app:sim} for the target states (a)~$\frac{1}{\sqrt{2}}(\ket{01}+\ket{10})$ and (b)~$\frac{1}{\sqrt{3}}\ket{01}+\frac{\sqrt{2}}{\sqrt{3}}\ket{10}$.
The results show that different objective functions yield similar numbers of tuning iterations; however, the QSV-based scheme significantly reduces both experimental and computational overheads.
Specifically, the QSV implementation requires only four measurement settings as compared with $16$ for QST, and the post-processing time per feedback round is reduced from $1.5\times10^{-4}$ s to $9.2\times10^{-6}$ s in case (a), and from $5.6\times10^{-3}$ s to $1.3\times10^{-5}$ s in case (b).

These simulations were performed prior to the experiments to guide the
choice of the learning rate used in the compensation algorithm. Rather
than scanning a predetermined interval systematically, we selected the
learning rate heuristically through preliminary simulations and pilot
tests, starting from an initial trial value and then adjusting it
according to the observed convergence behavior of the QSV passing
frequency. The final values adopted for the two target states shown in
Fig.~\ref{fig_app:sim} were $710$ for state (a) and $500$ for state (b),
chosen to maintain reliable convergence without
noticeable oscillation or instability during the prescriptive
preparation process.

\subsection{Results}

Following the approach described in the main text, we interpret the QSV procedure in the practical scenario as a hypothesis test. The verification outcome is then used to certify a fidelity lower bound, or equivalently an infidelity upper bound based on the test frequency and the spectral gap of the verification operator.
For fidelity estimation, we note that the optimal verification operator $\Omega_{\text{opt}}$ is not of homogeneous structure, and therefore does not support direct fidelity estimation via linear inversion. Instead, based on the applied protocol as in Eq.~\eqref{app_eq:QSV2}, we apply the general bounding method in Ref.~\cite{Zhu.Hayashi2019d} to obtain bounds on the estimated fidelity $F$ such that

\begin{align}
    \frac{1 - f}{\nu} \leq \epsilon_{\rm est} \leq 2\frac{1 - f}{1 - \nu}\,, \\
    1 - 2\frac{1 - f}{1 - \nu} \leq F_{\rm est} \leq \frac{f - (1 - \nu)}{\nu}\,,
\end{align}

where $f$ is the passing frequency and $\nu = 1/(2 + \sin\theta \cos\theta)$ denotes the spectral gap of the verification operator. 

\begin{figure*}[tb]
    \centering
    \includegraphics[width=\linewidth]{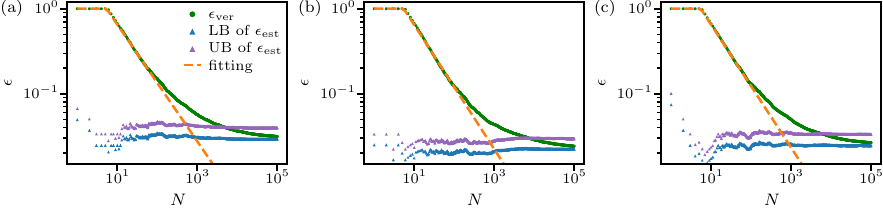}
    \caption{Verified infidelity $\epsilon_{\rm ver}$ (green dotted line) and bounds for estimated infidelity $\epsilon_{\rm est}$ (upper: purple triangles; lower: blue triangles) for the two-qubit states ${\ket{\psi(\theta)}=\sin\theta\ket{01}+\cos\theta\ket{10}}$ as in Eq.~\eqref{eq:twoqubittheta} with (a) ${\sin\theta = \frac{1}{\sqrt{3}}}$, (b) ${\sin\theta =\frac{1}{\sqrt{2}}}$, and (c) ${\sin\theta =\frac{\sqrt{2}}{\sqrt{3}}}$, as a function of the number of tests $N$. The significance level is set at $\delta < 0.05$, and all data points are obtained from 100 experimental trials. Dashed lines indicate the fitted scaling with $O(\epsilon^t)$, where $t = 1.45$, $1.37$, and $1.38$ respectively.}
    \label{FigS2}
\end{figure*}

Table~\ref{app_tab:2qubit_QSV} presents the certified fidelities ${F_{\rm ver} = 1-\epsilon_{\rm ver}}$ (significance level ${\delta=5\%}$) and estimated bounds across multiple values of $\theta$ and test numbers, highlighting the verification power of QSV even with limited samples. In particular, Fig.~\ref{FigS2} shows the behavior of verified and estimated infidelities versus the test number $N$ for representative states with ${\sin\theta = \left\{\frac{1}{\sqrt{3}}, \frac{1}{\sqrt{2}}, \frac{\sqrt{2}}{\sqrt{3}}\right\}}$, illustrating the scaling and statistical efficiency of the protocol.

\begin{table}[h]
\centering
\renewcommand{\arraystretch}{1.1} % 
\setlength{\tabcolsep}{10pt} % 
\begin{tabular}{l|ccccc}
\hline
$\boldsymbol{\sin\theta}$ & $0$ & $\frac{1}{\sqrt{3}}$ & $\frac{1}{\sqrt{2}}$ & $\frac{\sqrt{2}}{\sqrt{3}}$ & $1$ \\
\hline
Certified Fidelity ($N=20$)         & $64.6\%$ & $58.9\%$ & $60.6\%$ & $60.3\%$ & $62.9\%$ \\
Certified Fidelity ($N=50$)         & $84.2\%$ & $78.3\%$ & $80.5\%$ & $79.4\%$ & $82.5\%$ \\
Certified Fidelity ($N=10^2$)       & $91.0\%$ & $84.9\%$ & $87.8\%$ & $87.0\%$ & $89.3\%$ \\
Certified Fidelity ($N=10^4$)       & $99.0\%$ & $96.4\%$ & $97.1\%$ & $96.9\%$ & $98.3\%$ \\
Estimated Fidelity ($N=10^4$, lower bound)    & $99.0\%$ & $96.0\%$ & $97.0\%$ & $96.6\%$ & $98.3\%$ \\
Estimated Fidelity ($N=10^4$, upper bound)    & $99.3\%$ & $97.1\%$ & $97.8\%$ & $97.5\%$ & $98.7\%$ \\
\hline
\end{tabular}
\caption{Certified and estimated fidelities of the two-qubit states ${\ket{\psi(\theta)} = \sin\theta\ket{01} + \cos\theta\ket{10}}$ for various $\theta$ values. All values are derived from 100 experimental trials.}
\label{app_tab:2qubit_QSV}
\end{table}
%

%%%%%%
\section{Reconstruction fidelity and statistical analysis of the two-qubit states}\label{app:2qubitQST}
To benchmark the certification of the QSV strategies, we perform QST on these two-qubit states.
More significantly, we independently prepare these two-qubit states, tuning devices with the assistance of QST, to confirm the state preparation capability of the QSV method. 
In our QST experiments, each quantum state is measured in three independent runs to mitigate potential fluctuations caused by laser instability. Each run lasts for 1s, and the average coincidence counts from these measurements are used as the input frequency for simulation. In the simulation, we employ an efficient quantum state reconstruction algorithm~\cite{QSTshang} to evaluate the uncertainty. Specifically, 50 iterations are performed under various total photon number conditions to simulate the statistical uncertainty. For two-qubit states, the results indicate that the reconstruction uncertainty decreases progressively as the number of detected photons increases. When the total number of photons reaches around $10^6$, the standard deviation of the fidelity is reduced to approximately $10^{-4}$.

\begin{figure}[t]
    \centering
    \includegraphics[width=0.95\linewidth]{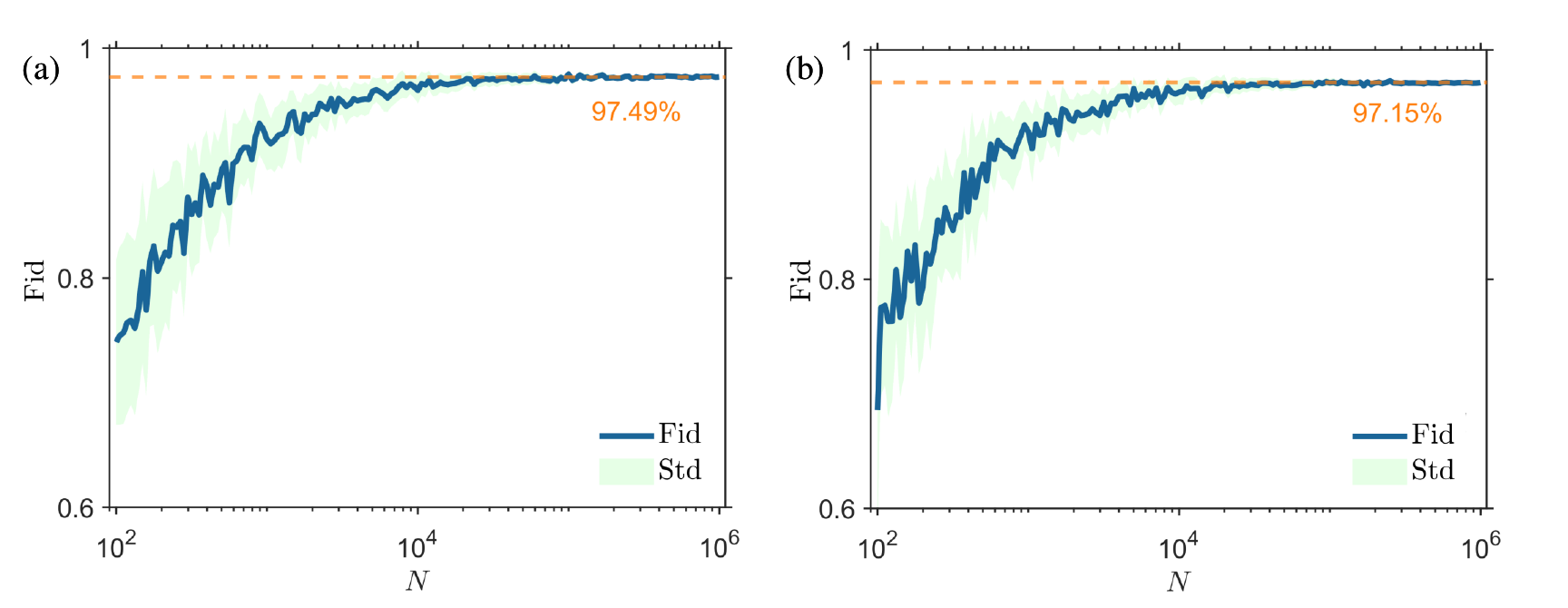}
    \caption{Fidelity (Fid) of the state $\frac{1}{\sqrt{2}}\bigl(\ket{01}_{\rm AB} + \ket{10}_{\rm AB}\bigr)$ reconstructed via QST, using (a) QSV and (b) QST as optimization indicators, respectively. In each panel, $N$ denotes the number of samples in a single round of the simulated experiment. The blue curve shows the average fidelity over 50 simulation rounds, while the shaded area indicates the standard deviation (Std). The orange dashed line corresponds to the fidelity obtained with a sufficiently large sample size (approximately $10^6$), serving as a benchmark for comparison.}

    \label{FigS3}
\end{figure}

We select the entangled state corresponding to ${\sin\theta = \frac{1}{\sqrt{2}}}$ to examine the relationship between fidelity and its standard deviation as a function of the photon number, as illustrated in Fig.~\ref{FigS3}. The simulation results demonstrate that as the number of detected photons increases, the fidelity of state reconstruction gradually improves, while its standard deviation decreases. Notably, when the photon count reaches around $10^6$, the fidelity converges.
Furthermore, by comparing the fidelity and standard deviation of the two-qubit states prepared using QSV and QST as optimization indicators, we observe that both methods yield final states of comparable quality. 
Results for all these two-qubit states are reported in Table~\ref{app_tab:2qubit_QST}.
This confirms that QSV, despite consuming much less quantum resource, is capable of serving as an effective fidelity indicator during the state preparation process.

\begin{table}[h]
    \centering
    \renewcommand{\arraystretch}{1.1} % 
    \setlength{\tabcolsep}{10pt} % 
    \begin{tabular}{c c c c c c}
    \toprule
        ~ & \multicolumn{2}{c}{\textbf{Prepared via QSV}}  & \multicolumn{2}{c}{\textbf{Prepared via QST}} & \textbf{Certification} \\ 
        \cmidrule(lr){2-3} \cmidrule(lr){4-5} \cmidrule(lr){6-6}
        $\boldsymbol{\sin\theta}$ & \textbf{Fid} & \textbf{Std ($10^{-4}$)} & \textbf{Fid} & \textbf{Std ($10^{-4}$)} & \textbf{Fid} ($\delta\leq5\%$) \\ 
    \midrule
        $0$                         & 99.59\% & 1.9 & 99.61\% & 1.5 & 99.0\% \\ 
        $\tfrac{1}{\sqrt{3}}$       & 97.84\% & 21  & 99.63\% & 8.8 & 96.4\% \\ 
        $\tfrac{1}{\sqrt{2}}$       & 97.49\% & 26  & 97.18\% & 24  & 97.1\% \\ 
        $\tfrac{\sqrt{2}}{\sqrt{3}}$& 98.56\% & 20  & 99.24\% & 7.2 & 96.9\% \\ 
        $1$                         & 99.56\% & 1.4 & 99.43\% & 1.5 & 98.3\% \\ 
    \bottomrule
    \end{tabular}
    \caption{The Fidelity (Fid) and standard deviation (Std) for the two-qubit states ${\ket{\psi(\theta)} = \sin\theta\ket{01} + \cos\theta\ket{10}}$ prepared by tuning devices via QSV and QST methods. The results are from QST protocols under 50 runs with each single run costing $10^{6}$ photons. The certified fidelity with a significant level $\delta<5\%$ via QSV using $10^4$ tests is also compared with (the same as in Table~\ref{app_tab:2qubit_QSV}).}
    \label{app_tab:2qubit_QST}
\end{table}

\section{Preparation and verification of $W_3$ state via QSV strategies}\label{app:W}
In our experiments, the three-qubit $W$ states are prepared on top of high-fidelity two-qubit entangled states (with a fidelity of 99.47\%) using a DOF multiplexing approach, as also described in the main text.
Briefly, by introducing a path qubit through polarization-path DOF multiplexing, photon~A is directed into a polarization-path encoding module consisting of three HWPs and two beam displacers (BD$1$ and BD$2$), as illustrated in Fig.~\ref{expset}(d).
This configuration introduces an additional path DOF for photon~A and may induce a random relative phase between the interferometers formed by BD$1$ and BD$2$.
After passing through the encoding module, the state is transformed into $\frac{1}{\sqrt{3}}(\ket{001}+\ket{100}+\ket{010})$, corresponding to the standard three-qubit $W_3$ state.
The tuning procedure that compensates for the random relative phase follows the same compensation scheme described above (Algorithm~\ref{alg:feedback}).

\begin{figure*}[h]
    \centering
    \includegraphics[width=0.65\linewidth]{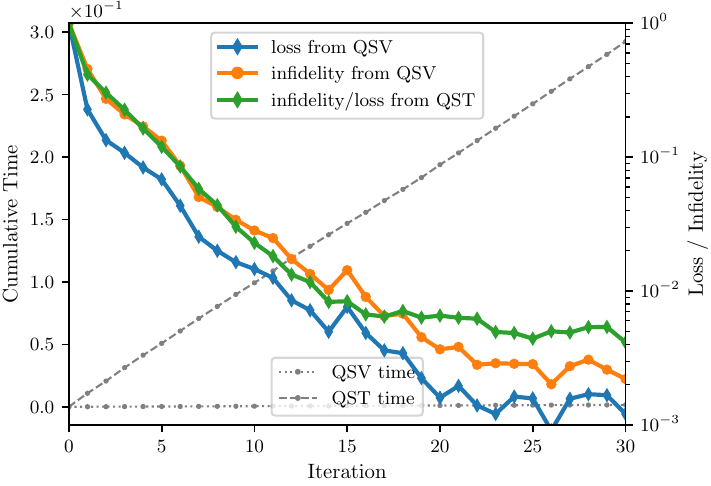}
    \caption{Simulation of prescriptive quantum-state preparation using the compensation algorithm for the three-qubit $W_3$ state.
    The post-processing time of the QSV-based method (gray dotted line) is significantly cheaper than QST (gray dashed line).
    The losses evaluated by different objective functions are compared: the passing frequency from QSV (blue diamonds) and the fidelity from full state tomography (green diamonds).
    The infidelity trajectory obtained from the QSV-based feedback is also shown (orange circles).
    The simulation starts from a noisy two-qubit state $\frac{1}{\sqrt{3}}\ket{01}+\frac{\sqrt{2}}{\sqrt{3}}\ket{10}$ with an initial fidelity of 99.47\%.
    All results are over 100 simulation rounds, and each data point is computed from a 1000-sample measurement.
    }
    \label{FigS4}
\end{figure*}

A simulation, shown in Fig.~\ref{FigS4}, was performed prior to the experiments to guide the choice of a suitable learning rate for the QSV-based indicator.
Following the same empirical selection strategy as
in the two-qubit case, we chose $\alpha=1.2$ for the three-qubit experiment, as it provides stable and sufficiently fast convergence of the QSV passing frequency.
For comparison, the simulation results obtained using a QST-based optimization are also included, demonstrating the efficiency and robustness of the QSV-based method.
Similarly, the QSV-based method has significantly reduced the post-processing time from $9.8\times10^{-3}$~s to $5.0\times10^{-5}$~s per round, achieving a fidelity comparable to that obtained with QST.

%

%%%%%%
\section{Reconstruction fidelity and statistical analysis of the $W_3$ state}\label{app:WQST}
To evaluate the performance of QSV in the preparation of multi-qubit entangled states, we extend our analysis to the three-qubit $W_3$ state. As a representative example, we first select a two-qubit polarization-entangled input state characterized by ${\sin\theta = \frac{1}{\sqrt{3}}}$, which provides a balanced entanglement structure suitable for its conversion into a $W_3$ state.

\begin{figure}[t]
    \centering
    \includegraphics[width=0.95\linewidth]{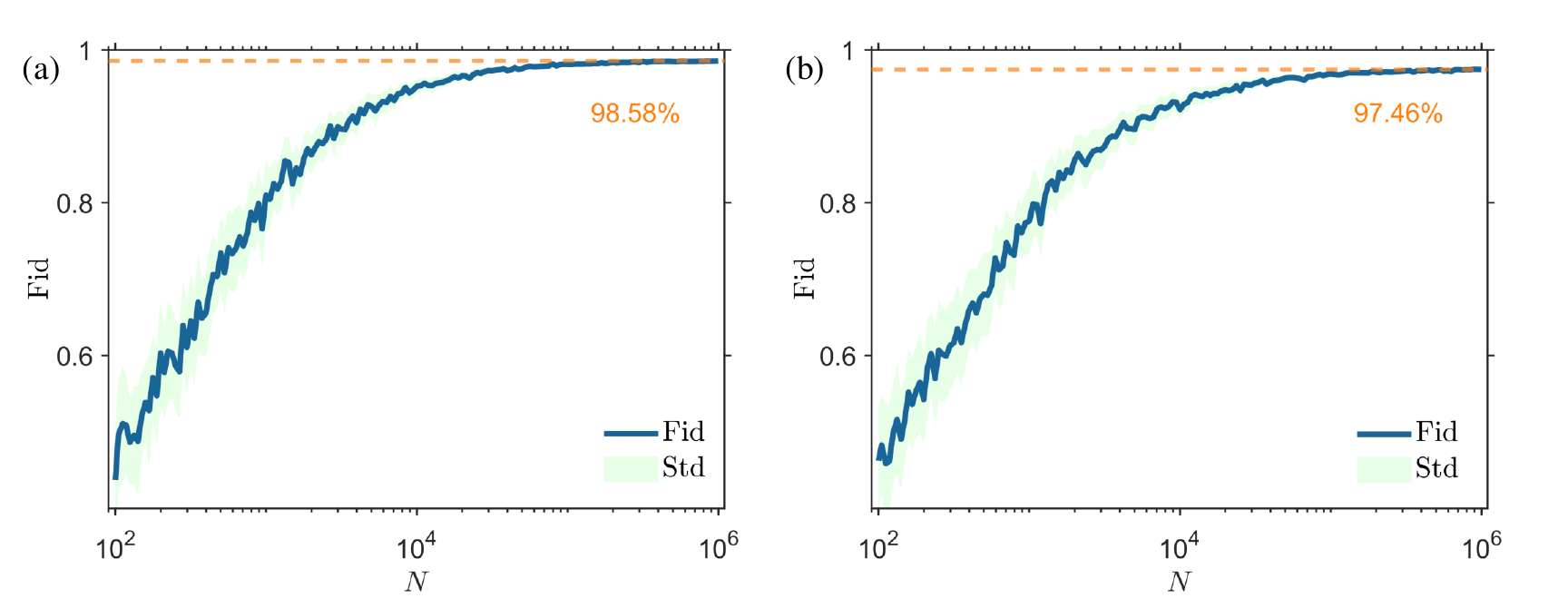}
    \caption{Fidelity (Fid) of the three-qubit $W_3$ state reconstructed via QST, using (a) QSV and (b) QST as optimization indicators, respectively. In each panel, $N$ denotes the number of samples in a single round of the simulated experiment. The blue curve shows the average fidelity over 50 simulation rounds, while the shaded area indicates the standard deviation (Std). The orange dashed line corresponds to the fidelity obtained with a sufficiently large sample size (approximately $10^6$), serving as a benchmark for comparison.}

    \label{FigS5}
\end{figure}

In the $W_3$ state preparation procedure, we adopt both QSV and QST as optimization indicators during the tuning of polarization-path DOF encoding module. After completing the optimization process, we carry out independent validation using both QSV and QST to assess the fidelity of the resulting three-qubit state.
Specifically, for QST-based analysis, we conduct simulated quantum state reconstruction in a manner analogous to the two-qubit case. The simulation involves varying the number of consumed photons per run to analyze the statistical convergence of the reconstructed fidelity. By iterating 50 trials for each photon number setting, we are able to characterize both the average fidelity and its corresponding standard deviation as functions of the resource consumption.
The results, as illustrated in Fig.~\ref{FigS5}, reveal that the reconstructed fidelities obtained using QSV- and QST-optimized states closely track each other across a wide range of photon number conditions. This consistency further demonstrates that QSV not only serves as a lightweight alternative to full QST, but also maintains high reliability and accuracy even in the more complex scenario of three-qubit entangled state preparation.

In summary, the successful extension of QSV-based optimization to the $W_3$ state confirms its feasibility and scalability. The close agreement between the QSV and QST results highlights the robustness of QSV for practical quantum information tasks that demand both speed and precision.

% %%%%
% \bibliography{QuantRefs}


\end{document}